\documentclass[thmsa]{article}
\usepackage{amsfonts}
\usepackage{amsmath}
\usepackage{amssymb}
\usepackage{amsfonts}
\usepackage{amsmath}
\usepackage{epsfig,multicol}
\usepackage{graphicx}
\usepackage{youngtab}

\input{tcilatex}

\begin{document}

\setcounter{page}{0} \topmargin0pt \oddsidemargin5mm \renewcommand{%
\thefootnote}{\fnsymbol{footnote}} \newpage \setcounter{page}{0}
\begin{titlepage}
\begin{flushright}
\end{flushright}
\vspace{0.5cm}
\begin{center}
{\Large {\bf PT-invariance and representations of the Temperley-Lieb algebra on the unit circle} }

\vspace{0.8cm}
{ \large Christian Korff}

\vspace{0.5cm}
{\em
Department of Mathematics, University of Glasgow, \\
15 University Gardens, Glasgow G12 8QW, UK}
\end{center}
\vspace{0.2cm}

\renewcommand{\thefootnote}{\arabic{footnote}}
\setcounter{footnote}{0}

\begin{abstract}
We present in detail a recent conjecture on self-adjoint representations of
the Temperley-Lieb algebra for particular values on the unit circle. The
formulation in terms of graphical calculus is emphasized and discussed for
several examples. The role of PT (parity and time reversal) invariance is
highlighted as it might prove important for generalizing the construction to
other cases.
\medskip
\par\noindent
\end{abstract}
\vfill{ \hspace*{-9mm}
\begin{tabular}{l}
\rule{6 cm}{0.05 mm}\\
c.korff@maths.gla.ac.uk
\end{tabular}}
\end{titlepage}\newpage

\section{Introduction}

We summarize recent novel results \cite{KW07,CK07} on the PT-invariant
construction of self-adjoint representations of the Temperley-Lieb algebra %
\cite{TL71}. The latter is defined as follows.

\begin{definition}
Let $\mathbb{C}(q)$ be the field of rational functions in an indeterminate $%
q $. The Temperley-Lieb algebra $TL_{N}(q)$ is the associative algebra over $%
\mathbb{C}(q)$ generated by $\{e_{1},...,e_{N-1}\}$ subject to the relations%
\begin{equation}
e_{i}^{2}=-(q+q^{-1})e_{i},\qquad e_{i}e_{i\pm 1}e_{i}=e_{i},\qquad
e_{i}e_{j}=e_{j}e_{i}\quad \text{for}\quad |i-j|>1~\text{.}
\label{TLrelations}
\end{equation}
\end{definition}

It is well-kown that if the indeterminate is evaluated
in the real numbers, $q\in \mathbb{R}$, or takes the special root-of-unity values
\begin{equation}
q=\exp (i\pi /r),\qquad r=3,4,5,...\;,  \label{pottsq}
\end{equation}
that there exist self-adjoint
representations of $TL_{N}(q)$ \cite{Jo83,We88,PPM91,Jo99}.
\smallskip

The novel aspect we wish to highlight here are recent results \cite{KW07} of
the construction of self-adjoint representations for the above values on the unit circle using ideas from quasi-Hermitian quantum mechanics and $PT$-invariance, e.g. \cite{SGH92,Most04,CFMFAF,Bender07}. This approach is physically motivated: it is a \emph{%
preparatory} step to explicitly construct for \emph{arbitrary} values $q\in $
$\mathbb{S}^{1}$ representations of $TL_{N}(q)$ for which the following
quantum Hamiltonian%
\begin{equation}
H=\sum_{i=1}^{N-1}e_{i}  \label{TLH}
\end{equation}%
is Hermitian or self-adjoint. The latter requirement is necessary in order
to allow for a sound physical interpretation of the associated quantum
model. Clearly, the case when each of the Temperley-Lieb generators $e_{i}$
is self-adjoint is a particular subset of this class of representations.
\smallskip

The case when $q$ is evaluated on the unit circle is of special physical
interest. For instance, at the aforementioned values (\ref{pottsq}) the
model is related to the $Q$-state Potts model with $Q=(q+q^{-1})^{2}$. For
the value $q=\exp (i\pi /2)$ the Hamiltonian (\ref{TLH}) describes critical
dense polymers on the square lattice, while for $q=\exp (2\pi i/3)$ it is
related to the problem of percolation. More generally, it has been argued%
\cite{PRZ06,RS07} that the Hamiltonian (\ref{TLH}) can be viewed as a
discrete system which either in the thermodynamic limit ($N\rightarrow \infty $) or through its algebraic properties
can be effectively described by logarithmic conformal field theories. While
these applications are beyond the scope of this article, it needs to be
stressed that they provide some of the basic motivation for the present
construction using quasi-Hermiticity and $PT$-invariance.
\smallskip

As we wish to outline the basic principles of the approach we shall focus on
the special case
\begin{equation}
q=\exp (i\pi /r),\qquad r>N,  \label{qspecial}
\end{equation}%
where $N$ is the number of strands and $r$ can take \emph{any real} values
greater than $N$ (not only integer values). This section of the unit circle
- while not of immediate physical interest as it shrinks to $q=1$ as $%
N\rightarrow \infty $ - is distinguished mathematically as it not only
allows for a self-adjoint representation of the Temperley-Lieb algebra but
also for the application of graphical calculus in terms of Kauffman diagrams
\cite{Kauf87}. This graphical formulation of the Temperley-Lieb algebra is
very elegant and greatly facilitates computations. For these reasons we wish
to maintain it when $q\in \mathbb{S}^{1}$. In this article we shall describe
in more detail a construction of an inner product \cite{CK07} which achieves
this for the values (\ref{qspecial}). Before we can start our discussion we
need to recall some previous results on PT-invariance and self-adjoint
representations of the Temperley-Lieb algebra.

\section{Review of previous results}

We will concentrate on the $U_{q}(sl_{2})$-invariant XXZ quantum spin-chain
model \cite{Alc87,PS90}. The latter model corresponds to a realisation of the
Hamiltonian (\ref{TLH}) in terms of the fundamental two-dimensional $%
U_{q}(sl_{2})$-module. Prior to introducing it, we recall the following
definition.

\begin{definition}
The q-deformed enveloping algebra (or quantum group) $U_{q}(sl_{2})\ $is the
associative algebra over $\mathbb{C}(q)$ generated by $\{E,F,K,K^{-1}\}$
subject to the relations%
\begin{equation}
KK^{-1}=K^{-1}K=1,\quad KEK^{-1}=q^{2}E,\quad KFK^{-1}=q^{-2}F,\quad \lbrack
E,F]=\frac{K-K^{-1}}{q-q^{-1}}~.
\end{equation}%
$U_{q}(sl_{2})$ can be endowed with structure of an Hopf algebra with
co-multiplication%
\begin{equation}
\Delta (K^{\pm 1})=K^{\pm 1}\otimes K^{\pm 1},\quad \Delta (E)=E\otimes
1+K\otimes E,\quad \Delta (F)=F\otimes K^{-1}+1\otimes F
\end{equation}%
and co-unit%
\begin{equation}
\varepsilon (E)=\varepsilon (F)=0,\qquad \varepsilon (K^{\pm 1})=1\ .
\end{equation}%
There is also an antipode but we will not use it in the following.
\end{definition}

Setting
\begin{equation}
V=\mathbb{C}v_{+}\oplus \mathbb{C}v_{-}
\end{equation}%
we define the two-dimensional fundamental $U_{q}(sl_{2})$-module by%
\begin{equation}
Ev_{+}=0,\qquad Ev_{-}=v_{+},\qquad Fv_{-}=0,\qquad Fv_{+}=v_{-},\qquad
Kv_{\pm }=q^{\pm 1}v_{\pm }\;.
\end{equation}%
There is a natural inner product on $V$ given by%
\begin{equation}
\langle v_{\sigma },v_{\sigma ^{\prime }}\rangle =\delta _{\sigma ,\sigma
^{\prime }},\qquad \sigma ,\sigma ^{\prime }=\pm 1\;.
\end{equation}%
We choose the inner product to be antilinear in the first factor. Consider now the $N$-fold tensor product $V^{\otimes N}$ of the fundamental
representation with the inner product%
\begin{equation}
\langle v_{\sigma _{1}}\otimes \cdots \otimes v_{\sigma _{N}},v_{\sigma
_{1}^{\prime }}\otimes \cdots \otimes v_{\sigma _{N}^{\prime }}\rangle
=\delta _{\sigma _{1},\sigma _{1}^{\prime }}\cdots \delta _{\sigma
_{N},\sigma _{N}^{\prime }}\;.  \label{origprod}
\end{equation}%
Then the Temperley-Lieb algebra has the following matrix representation over
$V^{\otimes N}$,%
\begin{equation}
e_{i}\mapsto \underset{i-1}{\underbrace{1\otimes \cdots 1}}\otimes
\boldsymbol{e}\otimes \underset{N-i-1}{\underbrace{1\cdots \otimes 1}}\;,
\label{TLrep}
\end{equation}%
where the matrix elements $\boldsymbol{e}_{\sigma ,\sigma ^{\prime
}}:=\langle v_{\sigma },\boldsymbol{e}v_{\sigma ^{\prime }}\rangle $ of the
operator $\boldsymbol{e}:V\otimes V\rightarrow V\otimes V$ are
\begin{equation}
(\boldsymbol{e}_{\sigma ,\sigma ^{\prime }})=\left(
\begin{array}{cccc}
0 & 0 & 0 & 0 \\
0 & -q^{-1} & 1 & 0 \\
0 & 1 & -q & 0 \\
0 & 0 & 0 & 0%
\end{array}%
\right) \ .  \label{TLrep2}
\end{equation}%
Note that with respect to the Hamiltonian (\ref{TLH}) the quantum group
provides a symmetry,%
\begin{equation*}
\lbrack H,U_{q}(sl_{2})]=0\text{\ .}
\end{equation*}%
This is a direct consequence of the quantum analogue of Schur-Weyl duality %
\cite{Jim86b}.
\smallskip

In the following we will now evaluate $q$ to be a complex number. If $q$ is
real then one finds that the above product is invariant with respect to the
action of the Temperley-Lieb algebra,%
\begin{equation}
q\in \mathbb{R}:\qquad \langle v,e_{i}w\rangle =\langle e_{i}v,w\rangle
,\quad v,w\in V^{\otimes N}\;.
\end{equation}%
For the quantum group generators one finds%
\begin{equation}
q\in \mathbb{R}:\qquad \langle v,\Delta ^{(N)}(x)w\rangle =\langle \Delta
_{op}^{(N)}(x)v,w\rangle ,\quad v,w\in V^{\otimes N},\;x\in U_{q}(sl_{2}),
\end{equation}%
where $\Delta _{op}=\tau \circ \Delta $ is the opposite coproduct with $\tau
$ being the "flip"-operator, $\tau (x\otimes y)=y\otimes x,$ and%
\begin{equation}
\Delta ^{(N)}=(1\otimes \Delta )\Delta ^{(N-1)}=(\Delta \otimes 1)\Delta
^{(N-1)},\qquad \Delta ^{(2)}\equiv \Delta \;.
\end{equation}%
The opposite coproduct $\Delta _{op}^{(N)}$ is defined analogously.
\smallskip

In contrast, if $q\neq \pm 1$ lies on the unit circle $\mathbb{S}^{1}$ the
inner product is no longer invariant,%
\begin{equation}
q\in \mathbb{S}^{1},\;q\neq \pm 1:\qquad \langle v,e_{i}w\rangle \neq
\langle e_{i}v,w\rangle ,\quad v,w\in V^{\otimes N}
\end{equation}%
and the Hamiltonian (\ref{TLH}) ceases to be Hermitian. On physical grounds
we therefore need to introduce a new inner product which renders $H$
Hermitian. For the values (\ref{pottsq}) and (\ref{qspecial}) it turns out
that this new inner product can also be chosen to be invariant with respect
to the Temperley-Lieb action \cite{KW07}. The language which we are going to
employ in the construction of the invariant product is physically motivated,
but as we will see the associated concepts have a clear mathematical
interpretation and can be generalized beyond the section (\ref{qspecial}) of
the unit circle.

\subsection{Quasi-Hermiticity and PT-invariance}

We wish to construct a map $\eta :V^{\otimes N}\rightarrow V^{\otimes N}$
which has the following properties:

\begin{enumerate}
\item It is Hermitian, $\langle v,\eta w\rangle =\langle \eta v,w\rangle $,
invertible, $\det \eta \neq 0$, and positive, $\eta >0$.

\item It intertwines the Hamiltonian and its Hermitian adjoint with respect
to the inner product (\ref{origprod}),%
\begin{equation}
\eta H=H^{\ast }\eta \;.  \label{Heta}
\end{equation}
\end{enumerate}

\noindent The properties listed uner (1) guarantee that the \emph{new }inner
product $\langle \cdot ~,\cdot ~\rangle _{\eta }:V^{\otimes N}\times
V^{\otimes N}\rightarrow \mathbb{C}$ given by%
\begin{equation}
\langle v,w\rangle _{\eta }:=\langle v,\eta w\rangle ,\qquad v,w\in
V^{\otimes N}  \label{etaprod}
\end{equation}%
is well-defined, while property (2) ensures that the Hamiltonian becomes
Hermitian,%
\begin{equation}
\langle Hv,w\rangle _{\eta }=\langle v,Hw\rangle _{\eta },\qquad v,w\in
V^{\otimes N}\;.  \label{Hherm}
\end{equation}%
Clearly, the existence of such a map $\eta $ is not guaranteed but needs to
be proved and - for all practical purposes - we wish to obtain $\eta $
explicitly. This has been achieved \cite{KW07} so far for (\ref{pottsq}) and
(\ref{qspecial}) using ideas from quantum group reduction \cite{RS90}. We
omit the details here and instead will explicitly state the new inner
product (and with it $\eta $) for (\ref{qspecial}) below, after introducing
a convenient graphical formalism.
\smallskip

Besides the above requirements, which render the Hamiltonian $H$
quasi-Hermitian, one can impose further constraints based on certain
transformation properties in connection with parity, time and spin-reversal.

\begin{definition}
Let $P$ (parity-reversal), $T$ (time-reversal) and $R$ (spin-reversal) be
the involutions $V^{\otimes N}\rightarrow V^{\otimes N}$ defined in terms of
the following action on the basis elements%
\begin{eqnarray*}
P &:&\;v_{\sigma _{1}}\otimes v_{\sigma _{2}}\cdots \otimes v_{\sigma
_{N}}\mapsto v_{\sigma _{N}}\otimes v_{\sigma _{N-1}}\cdots \otimes
v_{\sigma _{1}}, \\
T &:&\;\alpha ~v_{\sigma _{1}}\otimes \cdots \otimes v_{\sigma _{N}}\mapsto
\bar{\alpha}~v_{\sigma _{1}}\otimes \cdots \otimes v_{\sigma _{N}},\quad
\alpha \in \mathbb{C}, \\
R &:&\;v_{\sigma _{1}}\otimes \cdots \otimes v_{\sigma _{N}}\mapsto
v_{-\sigma _{1}}\otimes \cdots \otimes v_{-\sigma _{N}}\;.
\end{eqnarray*}%
We define $P,R$ to be linear but $T$ to be \emph{anti-linear}.
\end{definition}

\noindent \textbf{Remark}. While the definition of the parity and
spin-reversal operators is conceptually clear, the identification of the
antilinear map $T$ with time-reversal warrants an additional comment. One
important method to construct eigenvectors of the quantum Hamiltonian (\ref%
{TLH}) is the coordinate Bethe ansatz. The latter involves the definition of
a discrete quantum mechanical wave function $\psi $. The map $T$ is defined
such that this wave function is transformed into its complex conjugate, $%
\psi \rightarrow \bar{\psi}$. This transformation of the wave function which
ought to obey a discrete version of the Schr\"{o}dinger equation, $i\partial
_{t}\psi =H\psi $, corresponds to time-reversal.
\medskip

A straightforward computation exploiting (\ref{TLrep}), (\ref{TLrep2}) shows
that the Hamiltonian (\ref{TLH}) is $PT$ and $RT$-invariant. Namely, we have
that%
\begin{equation}
\langle Hv,w\rangle =\langle v,PHPw\rangle =\langle v,RHRw\rangle =\langle
v,THTw\rangle ,\qquad v,w\in V^{\otimes N}\ .  \label{PTRTH}
\end{equation}%
These identities motivate the following additioanl requirements on the map $%
\eta $,%
\begin{equation}
P\eta P=R\eta R=T\eta T=\eta ^{-1}\;.  \label{PTRTeta}
\end{equation}%
We shall refer to these transformations as $PT$ and $RT$-invariance of the
map $\eta $ and its associated inner product, respectively. An immediate
consequence of these relations is that%
\begin{equation}
\det \eta =\det P\eta ^{-1}P=\frac{1}{\det \eta }\;\Rightarrow \;\det \eta
=1\;.
\end{equation}%
As already mentioned such a map $\eta $ ensuring quasi-Hermiticity of the
Hamiltonian and satisfying $PT$ and $RT$-invariance does indeed exist and,
moreover, can be explicitly constructed \cite{KW07}. We summarize the
previous results \cite{KW07} in the following theorem.

\begin{theorem}
Evaluate $q$ in the section (\ref{qspecial}) of the unit circle. Then there
exists a map $\eta :V^{\otimes N}\rightarrow V^{\otimes N}$ possessing the
properties mentioned above and in addition enjoys the more restrictive
constraints%
\begin{equation}
\langle e_{i}v,w\rangle _{\eta }=\langle v,e_{i}w\rangle _{\eta },\qquad
i=1,2,...,N-1  \label{TLinv}
\end{equation}%
and%
\begin{equation}
\langle \Delta _{op}^{(N)}(\varphi (x))v,w\rangle _{\eta }=\langle v,\Delta
^{(N)}(x)w\rangle _{\eta },\qquad x\in U_{q}(sl_{2})  \label{QGinv}
\end{equation}%
where $v,w\in V^{\otimes N}$ and $\varphi $ is the $U_{q}(sl_{2})$%
-automorphism%
\begin{equation}
\varphi (K^{\pm 1})=K^{\mp 1},\quad \varphi (E)=F,\quad \varphi
(F)=E,\quad \varphi (xy)=\varphi (y)\varphi (x),\; x,y\in
U_{q}(sl_{2})\;.
\end{equation}
\end{theorem}

If one explicitly computes $\eta $ using the previous results in the
literature \cite{KW07}, one quickly realizes the importance of the choice of
basis. For example, if $\eta $ is to be computed with respect to the basis
vectors%
\begin{equation}
\{v_{\sigma _{1}}\otimes \cdots \otimes v_{\sigma _{N}}~|~\sigma _{i}=\pm
1\}\subset V^{\otimes N}  \label{spinbasis}
\end{equation}%
one finds in general none of its matrix element are nonzero within a fixed
spin sector, i.e. in a subspace with $\sum_{i}\sigma _{i}=~$const. This
makes an evaluation of the inner product (\ref{etaprod}) and computations of
physically relevant quantities extremely cumbersome. We therefore shall
introduce another basis which is algebraically motivated: it transforms very
simply under the action of the Temperley-Lieb algebra and the quantum group.
Most importantly, in this basis the action of the latter two algebras can be
described graphically.

\section{Change of basis and graphical calculus}

The new basis, we denote it by $\{t_{i}\}$, is not \emph{orthonormal} and
leads to a shift in emphasis from the map $\eta :V^{\otimes N}\rightarrow
V^{\otimes N}$ discussed above to the Gram matrix%
\begin{equation}
G_{ij}=\langle t_{i},\eta t_{j}\rangle \ .  \label{G}
\end{equation}%
The latter nicely reflects the algebraic properties of the new basis and can
be evaluated by graphical means. Crucial for this graphical evaluation is
the fact that there is a correspondence between basis vectors $t_{i}$ and
elements $a_{i}$ in the Temperley-Lieb algebra. Using this correspondence
one can directly define the Gram matrix in terms of a real functional
\begin{equation}
\omega :TL_{N}(q)\rightarrow \mathbb{R}
\end{equation}%
by setting%
\begin{equation}
G_{ij}:=\omega (a_{i}^{\ast }a_{j})  \label{Gom}
\end{equation}%
where $\ast $ denotes a conjugation in the Temperley--Lieb algebra which
corresponds to taking the Hermitian adjoint in the associated
representation. The values of the functional can be computed using Kauffman
diagrams. This suggests to circumvent the construction of the map $\eta $ in
the spin basis (\ref{spinbasis}) entirely and instead to focus on the
algebraically distinguished Gram matrix $G$. Obviously, all the properties
of $\eta $ can be translated to properties of the matrix $G$ and it is
convenient to do so because many matrix elements of $G$ turn out to be
vanishing. We shall list the properties of the Gram matrix $G$ below. First
we introduce the new basis vectors.

\subsection{The new basis in terms of Young tableaux}

The new basis $\{t_{i}\}$ which are going to define is closely related to
the dual canonical basis discussed by Frenkel and Khovanov \cite{FK}, see also \cite{Lu} and \cite{CS} and references therein. The
alert reader will notice, however, that there are important differences in
our conventions from the ones used by the latter authors, since we need to
accommodate that $q$ lies on the unit circle. In particular our definition
of the inner product, respectively the Gram matrix $G$, differs from the one
for $q$ real where the construction of Frenkel and Khovanov applies.
\smallskip

We start by decomposing the representation space $V^{\otimes N}$ with
respect to the number of "down spins",%
\begin{equation}
V^{\otimes N}=\tbigoplus_{n=0}^{N}W_{n},\qquad W_{n}=\limfunc{span}_{\mathbb{%
C}}\left\{ v_{\sigma _{1}}\otimes \cdots \otimes v_{\sigma
_{N}}~|~\tsum\nolimits_{i}\sigma _{i}=N-2n\right\} \ .  \label{spindecomp}
\end{equation}%
Note that the action of the Temperley-Lieb algebra respects this
decomposition,%
\begin{equation}
TL_{N}(q)W_{n}=W_{n}\;.
\end{equation}%
In each subspace $W_{n}$ we now introduce the following basis \cite{Stroppel}. Let $\lambda
_{n}$ be the rectangular Young diagram with $n$ rows of $N-n$ boxes,%
\begin{equation*}
\lambda _{n}=\underset{N-n}{\underbrace{\left.
\begin{tabular}{|l|l|l|l|l|}
\hline
&  &  &  &  \\ \hline
&  &  &  &  \\ \hline
&  &  &  &  \\ \hline
&  &  &  &  \\ \hline
\end{tabular}%
\right\} }}~n
\end{equation*}%
Then we assign to each subdiagram $\lambda ^{\prime }\subset \lambda _{n}$ a
vector in $W_{n}$ as follows. Let $t$ be the unique standard tableau (column
and row strict) of shape $\lambda ^{\prime }$ whose entries are consecutive
integers with entry $n$ in the upper left corner. For example,%
\begin{equation}
t=%
\begin{tabular}{|c|cccc}
\hline
$n$ & $n+1$ & \multicolumn{1}{|c}{$n+2$} & \multicolumn{1}{|c}{$\cdots $} &
\multicolumn{1}{|c|}{$s$} \\ \hline
$n-1$ & $n$ & \multicolumn{1}{|c}{$\cdots $} & \multicolumn{1}{|c}{$s-2$} &
\multicolumn{1}{|c}{} \\ \cline{1-4}
$\vdots $ &  & \multicolumn{1}{|c}{} & \multicolumn{1}{|c}{} &  \\
\cline{1-3}
$s^{\prime }$ &  &  &  &  \\ \cline{1-1}
\end{tabular}%
~,\quad n<s<N,\quad 1\leq s^{\prime }<n\;.
\end{equation}%
Reading the entries of the tableau from left to right and top to bottom we
set%
\begin{equation}
t\mapsto e_{s^{\prime }}e_{s^{\prime }-1}\cdots e_{s-2}\cdots
e_{n-1}e_{s}\cdots e_{n+1}e_{n}\Omega _{n},
\end{equation}%
where
\begin{equation}
\Omega _{n}=\underset{n}{\underbrace{v_{-}\otimes v_{-}\cdots \otimes v_{-}}}%
\otimes v_{+}\otimes v_{+}\cdots \otimes v_{+}
\end{equation}%
is the vector corresponding to $\lambda ^{\prime }=\varnothing $. Note that
for fixed $n$ there are as many of these tableaux as the dimension of the
subspace $W_{n}$, namely $\dim W_{n}=\dbinom{N}{n}$.
\smallskip

\noindent \textbf{Example}. Let $N=5$ and $n=2$ then we have the following Young
diagrams and tableaux:

\begin{equation*}
\Yvcentermath1t=\varnothing,\quad\young(2),\quad \young(2,1),\quad \young(23),\quad \young%
(23,1),\quad \young(23,12),\quad \young(234),\quad \young(234,1),\quad \young%
(234,12),\quad \young(234,123)\;.
\end{equation*}%
The corresponding algebra elements $a\in TL_{N}(q)$ are
\begin{equation*}
a=1,\; e_{2},\; e_{1}e_{2},\; e_{3}e_{2},\; e_{1}e_{3}e_{2},\;
e_{2}e_{1}e_{3}e_{2},\; e_{4}e_{3}e_{2},\; e_{1}e_{4}e_{3}e_{2},\;
e_{2}e_{1}e_{4}e_{3}e_{2},\; e_{3}e_{2}e_{1}e_{4}e_{3}e_{2}\ .
\end{equation*}

Note that we do not distinguish in our notation between Young tableaux and
the associated basis vectors. Henceforth, we will also often identify the
basis vectors respectively tableaux with the corresponding algebra elements.

\subsection{Kauffman and oriented cup diagrams}

There is an elegant graphical calculus connected with the new basis.
Represent each algebra element in $TL_{N}(q)$ in terms of Kauffman diagrams,
see the graphical depiction below,%
\begin{equation*}
\includegraphics[scale=0.6]{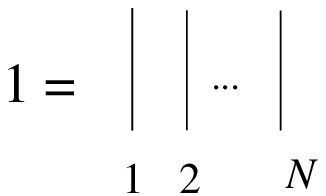}\qquad \text{and}\qquad %
\includegraphics[scale=0.6]{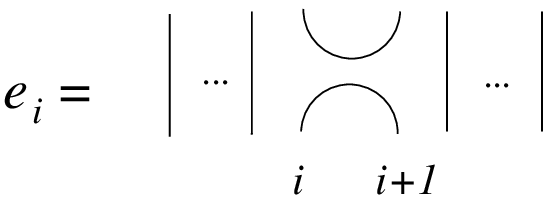}
\end{equation*}%
and realize the multiplication through concatenation from above. To compute
the action of the algebra on the basis $\{t_{i}\}$ we can identify each
basis vector with a half or cup-diagram which carries an orientation.
Inspired by the formalism of Frenkel and Khovanov \cite{FK} we introduce the
following graphical rules. Define a "cap" to be the map $\cap :V\otimes
V\rightarrow \mathbb{C}$ with%
\begin{equation}
v_{+}\otimes v_{+}\mapsto 0,\qquad v_{+}\otimes v_{-}\mapsto -q^{-1},\qquad
v_{-}\otimes v_{+}\mapsto 1,\qquad v_{-}\otimes v_{-}\mapsto 0
\end{equation}%
and a "cup" $\cup $\ to be the map $\cup :\mathbb{C}\rightarrow V\otimes V$
such that%
\begin{equation}
1\mapsto v_{+}\otimes v_{-}-q~v_{-}\otimes v_{+}\ .
\end{equation}%
Graphically these maps are represented as follows%
\begin{equation*}
\includegraphics[scale=0.6]{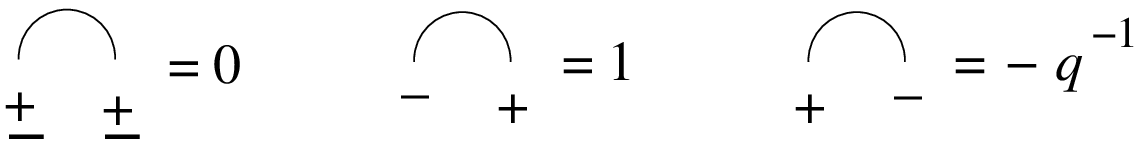}
\end{equation*}%
and%
\begin{equation*}
\includegraphics[scale=0.6]{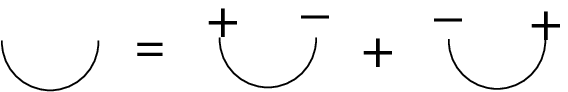},\qquad
\includegraphics[scale=0.6]{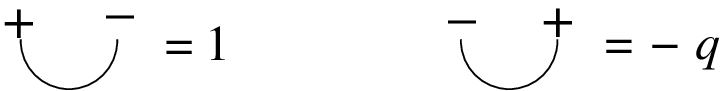}
\end{equation*}%
Each Temperley-Lieb generator $e_{i}:V_{i}\otimes V_{i+1}\rightarrow
V_{i}\otimes V_{i+1}$ then corresponds to the composition
\begin{equation}
e_{i}=\cup _{i,i+1}\circ \cap _{i,i+1},
\end{equation}%
where $V_{i}\otimes V_{i+1}$ are the $i^{\text{th}}$ and ($i+1$)$^{\text{th}%
} $ copy in the tensor product $V^{\otimes N}$. Employing the graphical rules%
\begin{equation*}
\includegraphics[scale=0.6]{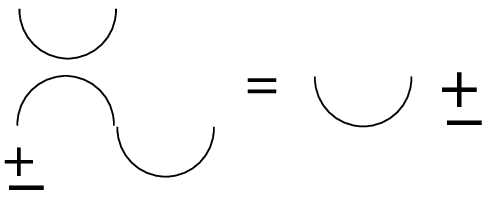}\qquad
\includegraphics[scale=0.6]{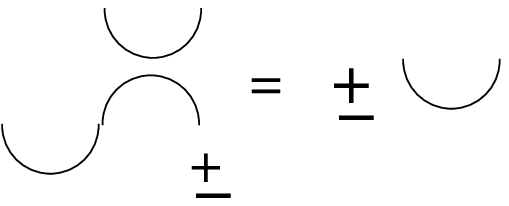}
\end{equation*}%
and
\begin{equation*}
\includegraphics[scale=0.6]{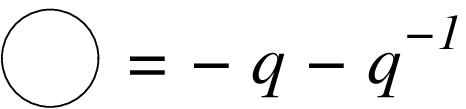},
\end{equation*}%
one can now easily generate the basis $\{t_{i}\}$ by acting with the
corresponding algebra elements $\{a_{i}\}$ on the cyclic vector $\Omega _{n}$%
. Applying the same graphical
rules one also easily deduces that the action of $TL_{N}(q)$ simply permutes
the basis elements $\{t_{i}\}$ up to factors of $-(q+q^{-1})$.
\smallskip

Besides a simple action of the Temperley-Lieb algebra the new basis vectors
display also a nice transformation behaviour under the action of the quantum
group \cite{FK}. Notice that the spin sectors are not preserved under the $%
U_{q}(sl_{2})$-action, instead we have for the quantum group generators $E,F$
that
\begin{equation}
E:W_{n}\rightarrow W_{n-1}\qquad \text{and}\qquad F:W_{n}\rightarrow
W_{n+1}\;.
\end{equation}%
We now describe the action of $E$. Suppose we are given a cup diagram/Young
tableaux $t$ with $k_{-}$ down spins (minus signs) then%
\begin{equation}
E=\sum_{m=1}^{k_{-}}[m]_{q}E_{m},\qquad \lbrack m]_{q}:=\frac{q^{m}-q^{-m}}{%
q-q^{-1}},
\end{equation}%
where $E_{m}$ connects the $m^{\text{th}}$ and ($m+1$)$^{\text{th}}$ down
spin with a cup (here any intermediate cups are ignored in the counting).
For $m=k_{-}$ the map $E_{k}$ simply flips the rightmost down-spin (minus
sign) to an up-spin (plus sign). Note that by construction of the basis $%
\{t_{i}\}$ all down-spins are to the left of all up-spins. Similarly, the
action of $F$ on a cup diagram with $k_{+}$ up-spins can be described in
terms of the sum%
\begin{equation}
F=\sum_{m=1}^{k_{+}}[m]_{q}F_{m},
\end{equation}%
where $F_{m}$ connects the $m^{\text{th}}$ and ($m+1$)$^{\text{th}}$ up-spin
and one starts counting from the right. Again, any intermediate cups are
ignored and $F_{k_{+}}$ simply flips the leftmost up-spin to a down spin.
\medskip

\noindent \textbf{Example}. Let $N=5$ and take the tableau%
\begin{equation*}
\Yvcentermath1t=\young(2,1)=+\cup ++\;.
\end{equation*}%
Then we have $k_{+}=3$ and%
\begin{equation*}
F~+\cup ++=(+\cup \cup )+[2]_{q}(\Cup +)+[3]_{q}(-\cup ++)\;.
\end{equation*}

Using the above graphical calculus one can now easily verify that the
matrices defined through%
\begin{equation}
e_{i}t_{j}=\sum_{k}t_{k}(\varepsilon _{i})_{kj}  \label{TLmatrix}
\end{equation}%
and%
\begin{equation}
Et_{j}=\sum_{k}t_{k}\mathcal{E}_{kj},\qquad Ft_{j}=\sum_{k}t_{k}\mathcal{F}%
_{kj}  \label{EFmatrix}
\end{equation}%
are all real valued, but not symmetric. Because of (\ref{TLmatrix}) this
particularly applies also to the matrix of the Hamiltonian (\ref{TLH}),%
\begin{equation}
Ht_{j}=\sum_{k}t_{k}\mathcal{H}_{kj},\qquad \mathcal{H}_{kj}\in \mathbb{R}%
\qquad \text{and}\qquad \mathcal{H}\neq \mathcal{H}^{t}\;.  \label{Hmatrix}
\end{equation}%
In contrast, the Hamiltonian matrix with respect to the spin basis (\ref%
{spinbasis}) is symmetric but not real valued. Thus, as before we need to
introduce a new inner product in terms of the Gram matrix (\ref{G}) with
respect to which the Hamiltonian becomes Hermitian.

\section{Graphical definition of the Gram matrix}

We are now ready to introduce the Gram matrix with respect to the new basis
defined in the previous section. Since we wish to generalize the
construction described here to other values of $q$ on the unit circle in
future work (see the comments in the introduction), it is worthwhile to
first formulate its general properties before specializing to the section (%
\ref{qspecial}).
\smallskip

The analogues of property (1) and (2) for $\eta $ are:

\begin{enumerate}
\item $G$ is Hermitian, $G_{ij}=\bar{G}_{ji}$, invertible, $\det G\neq 0,$
and positive, $G>0$.

\item $G$ intertwines the matrix $\mathcal{H}$ with its transpose,%
\begin{equation}
G\mathcal{H=H}^{t}G\;.
\end{equation}
\end{enumerate}

\noindent Besides these "minimal" requirements on $G$ we can impose the
additional constraints originating from PT and RT-invariance. Namely, from
the equalities (\ref{PTRTeta}) we deduce that%
\begin{equation*}
G_{ij}=\langle Tt_{i},\eta ^{-1}Tt_{j}\rangle =\langle PTt_{i},\eta
PTt_{j}\rangle =\langle RTt_{i},\eta RTt_{j}\rangle
\end{equation*}%
and hence%
\begin{equation*}
\pi ^{\ast }G\pi =G\qquad \text{and}\qquad \rho ^{\ast }G\rho =G,
\end{equation*}%
where
\begin{equation*}
PT~t_{i}=\sum_{j}t_{j}\pi _{ji}\qquad \text{and}\qquad
RT~t_{i}=\sum_{j}t_{j}\rho _{ji}\;.
\end{equation*}%
Employing Hermiticity in conjunction with time-reversal, we find that the
matrix%
\begin{equation*}
\eta _{\boldsymbol{\sigma },\boldsymbol{\sigma }^{\prime }}=\langle
v_{\sigma _{1}}\otimes \cdots \otimes v_{\sigma _{N}},\eta v_{\sigma
_{1}^{\prime }}\otimes \cdots \otimes v_{\sigma _{N}^{\prime }}\rangle
\end{equation*}%
obeys the identities $\eta ^{-1}=\eta ^{t}=\bar{\eta}=T\eta T$ from which we
conclude that%
\begin{equation}
G_{ij}=G_{ji}\in \mathbb{R}\ .
\end{equation}%
Finally, $G$ also inherits the following properties from $\eta $: $G$ is
block-diagonal with respect to the decomposition (\ref{spindecomp}) and in
addition $\det G=1$.
\smallskip

We now introduce for (\ref{qspecial}) a Gram matrix which satisfies all of
these requirements and, furthermore, is invariant under the action of the
Temperley-Lieb algebra. As already hinted at previously we define $G$ in
terms of a functional over the Temperley-Lieb algebra which can be computed
in terms of Kauffman diagrams. This functional is the subject of the next
definition.

\begin{definition}
Identify each $a\in TL_{N}(q)$ with its Kauffman diagram and fix an integer $%
0\leq n\leq N$. Assign to the top and bottom of the diagram the orientation
\begin{equation*}
\boldsymbol{\sigma }_{n}=\{\underset{n}{\underbrace{-,...,-}},\underset{N-n}{%
\underbrace{+,...,+}}\}\;.
\end{equation*}%
Let $x$ be the number of anti-clockwise oriented cups%
\begin{equation*}
\includegraphics[scale=0.6]{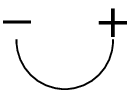},
\end{equation*}%
$y$ the number of closed loops, and $z$ the number of unoriented cups, caps
or through lines,%
\begin{equation*}
\includegraphics[scale=0.6]{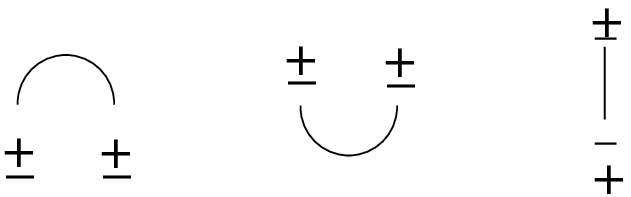}.
\end{equation*}%
Then we define the following functional $\omega _{n}:TL_{N}(q)\rightarrow
\mathbb{R}$ by setting\footnote{%
Previously, we distinguished the case $N$ odd and even \cite{CK07}. However,
simplifying the expression for $N$ odd one can see that both cases are
described by the same formula.}%
\begin{equation*}
a\mapsto \omega _{n}(a)=\left\{
\begin{array}{cc}
(-)^{x+y}(q+q^{-1})^{y}\dfrac{q^{\frac{N}{2}-n}+q^{n-\frac{N}{2}}}{q^{\frac{N%
}{2}-x}+q^{x-\frac{N}{2}}}, & \text{if }z=0 \\
0, & \text{else}%
\end{array}%
\right. ~.
\end{equation*}
\end{definition}

\noindent \textbf{Example}. We illustrate the above definition for two
examples. Let $N=5,~n=2$ and set $a=e_{1}e_{2}e_{3}e_{2}$ and $%
b=e_{3}e_{2}e_{1}e_{4}e_{3}e_{2}$. Then the associated oriented Kauffman
diagrams are%
\begin{equation*}
\includegraphics[scale=0.8]{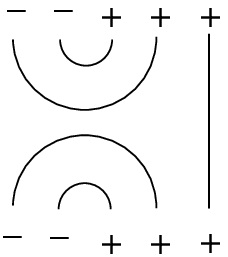}\qquad\text{and}\qquad
\includegraphics[scale=0.8]{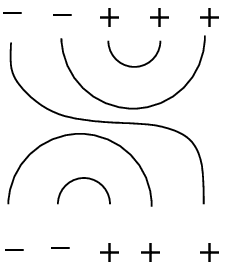}\;.
\end{equation*}%
We thus obtain%
\begin{equation*}
\omega _{n=2}(a)=1\qquad \text{and}\qquad \omega _{n=2}(b)=0\;.
\end{equation*}

\begin{conjecture}
Let $\{t_{i}\}$ denote the basis of $V^{\otimes N}$ described above in terms
of Young tableaux and $\{a_{i}\}\subset TL_{N}(q)$ be the corresponding
algebra elements. For each $n=0,1,2,...,N$ we set%
\begin{equation}
G_{ij}=\langle t_{i},\eta t_{j}\rangle :=\omega _{n}(a_{i}^{\ast
}a_{j}),\qquad \forall t_{i},~t_{j}\in W_{n}\ .
\end{equation}%
Here $\ast :TL_{N}(q)\rightarrow TL_{N}(q)$ is the antilinear automorphism
defined by%
\begin{equation}
(e_{i_{1}}e_{i_{2}}\cdots e_{i_{k}})^{\ast }=e_{i_{k}}e_{i_{k-1}}\cdots
e_{i_{1}}
\end{equation}%
and in terms of Kauffman diagrams is realized by flipping at the horizontal
axis. The matrix $G$ satisfies all of the above identities, in particular
those arising from $PT$-invariance, and in addition obeys the relations%
\begin{equation}
G\mathcal{E}=\mathcal{F}^{t}G\qquad \text{and}\qquad G\varepsilon
_{i}=\varepsilon _{i}^{t}G,\quad \quad i=1,...,N-1,
\end{equation}%
where $\mathcal{E},\mathcal{F}$ and $\varepsilon _{i}$ are the matrix
expressions for the quantum group and Temperley-Lieb generators in the basis
$\{t_{i}\}$ as introduced earlier, see (\ref{EFmatrix}) and (\ref{TLmatrix}).
\end{conjecture}

\noindent \textbf{Remark}. The implicit definition of the map $\eta $
contained in the above expression for the Gram matrix $G$ coincides with the
earlier construction \cite{KW07}. In fact, this is one way of checking the
above conjecture \cite{CK07}. Alternatively, one can verify the various
identities and properties independently of the map $\eta $ and this the
point of view which we have taken here. Numerical checks of the above
conjecture have been carried out for $N=2,3,4,5,6,7,8$.
\medskip

Note that there are simplifications in the computation of the inner product
if we restrict to certain subspaces $W^{\prime }$ which are left invariant
under the Temperley-Lieb action within a fixed sector $W_{n}$. Namely, set $%
n=\left\lfloor N/2\right\rfloor $ (the integer part of $N/2$) and consider
the subspace $W_{\max }\subset W_{n}$ of cup diagrams with a maximal number
of cups. From the graphical calculus reviewed earlier, it is clear that
\begin{equation*}
TL_{N}(q)W_{\max }=W_{\max }
\end{equation*}%
and that for $N$ even%
\begin{equation*}
EW_{\max }=FW_{\max }=\{0\}\ .
\end{equation*}%
If $N$ is odd we have obviously two subspaces $W_{\max }^{\pm }$ due to one
unpaired vector or spin with
\begin{equation*}
EW_{\max }^{-}=W_{\max }^{+}\qquad \text{and}\qquad FW_{\max }^{+}=W_{\max
}^{-}\;.
\end{equation*}%
In terms of Young tableaux these subspaces are spanned by all $t$ which
contain the "staircase" tableaux%
\begin{equation*}
\begin{tabular}{|c|cccc}
\hline
$n$ & $n+1$ & \multicolumn{1}{|c}{$\cdots $} & \multicolumn{1}{|c}{$\cdots $}
& \multicolumn{1}{|c|}{$\overset{}{\underset{}{2\left\lfloor \frac{N}{2}%
\right\rfloor -1}}$} \\ \hline
$\underset{}{\overset{}{n-1}}$ & $n$ & \multicolumn{1}{|c}{$\cdots $} &
\multicolumn{1}{|c}{} & \multicolumn{1}{|c}{} \\ \cline{1-4}
$\vdots $ &  & \multicolumn{1}{|c}{} & \multicolumn{1}{|c}{} &  \\
\cline{1-3}
$\overset{}{\underset{}{2}}$ & $1$ & \multicolumn{1}{|c}{} &  &  \\
\cline{1-2}
$\underset{}{\overset{}{1}}$ &  &  &  &  \\ \cline{1-1}
\end{tabular}%
\ .
\end{equation*}%
The corresponding Kauffman diagrams for $N$ even and odd look as follows,%
\begin{equation*}
\includegraphics[scale=0.6]{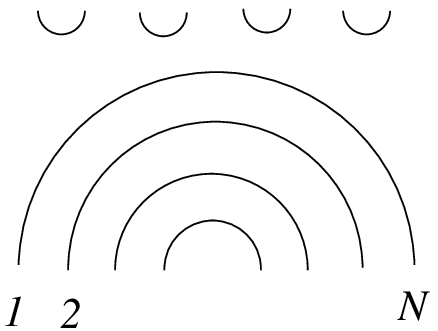}\qquad \text{and}\qquad
\includegraphics[scale=0.6]{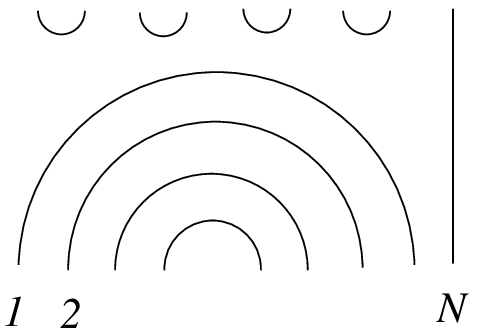}\;.
\end{equation*}%
From this graphical representation it is apparent that for each Kauffman
diagram associated with the algebra element $a_{i}^{\ast
}a_{j},\;a_{i},a_{j}\in W_{\max }$ the number $x$ of anti-clockwise oriented
cups equals the number $n$ of down spins (minus signs). Thus, we may for
this special case compute the Gram matrix purely in terms of the upper half
diagrams (= cup diagrams) as we only need to count the number of closed
loops, i.e.%
\begin{equation*}
\left( G|_{W_{\max }}\right) _{ij}=\omega _{n}(a_{i}^{\ast
}a_{j})=(-q-q^{-1})^{y_{ij}},
\end{equation*}%
where $y_{ij}=y_{ji}$ is the number of closed loops in $a_{i}^{\ast }a_{j}$.
We illustrate the comments just made for a simple example.
\smallskip

\noindent \textbf{Example}. Set $N=7$ and $n=3$. Then the subspace $W_{\max
} $ of all half-diagrams containing 3 cups is spanned by the tableaux $t$
which contain
\begin{equation*}
\Yvcentermath1\young(345,23,1)~\equiv \cup \cup \cup +\quad .
\end{equation*}%
Here we have expressed the corresponding basis vector
\begin{equation*}
e_{1}e_{3}e_{2}e_{5}e_{4}e_{3}~v_{-}\otimes v_{-}\otimes v_{-}\otimes
v_{+}\otimes v_{+}\otimes v_{+}\otimes v_{+}
\end{equation*}%
in terms of a cup diagram by omitting the lower half diagram. If we now wish
to compute the scalar product between the two diagrams%
\begin{equation*}
\Yvcentermath1\young(345,23,1)~\equiv \cup \cup \cup +\quad \text{and}\quad %
\Yvcentermath1\young(3456,23,1)~\equiv \cup \cup +\cup \text{\ }
\end{equation*}%
we simply need to combine them by flipping one of them at the horizontal
axis and count the number of closed loops. We find
\begin{equation*}
\langle \cup \cup \cup +,\cup \cup +\cup \rangle =(q+q^{-1})^{2}\;.
\end{equation*}%
\textbf{Note}, that this simplification of computing the Gram matrix purely
in terms of half or cup-diagrams is not possible in general when computing
scalar products between diagrams which differ in their number of cups. For
instance, to determine the scalar product between a vector in $W_{n}$ and
the cyclic vector $\Omega _{n}$ requires the full diagram. The computation
of these scalar products is necessary to ensure quasi-Hermiticity of the
Hamiltonian on the entire state space $V^{\otimes N}$.

\section{Discussion}

The Temperley-Lieb algebra arises in the context of quantum integrable
models whose dynamics and quantum statistics is described by the Hamiltonian
(\ref{TLH}) and we have discussed for a special example how to render it
Hermitian by constructing an appropriate inner product. These quantum
integrable models are closely related to classical two-dimensional
statistical mechanics models which are defined in terms of the following
solution of the quantum Yang-Baxter equation,%
\begin{equation}
R_{i,i+1}(x)=\boldsymbol{1}+\frac{x-x^{-1}}{xq-x^{-1}q^{-1}}~e_{i}\ .
\end{equation}%
There are various integrable boundary conditions one can impose on a
square-lattice and here we only concentrated on those which turn the model
quantum group invariant. In context of the six-vertex model solution the
statistical transfer matrix then reads
\begin{equation*}
t(x)=\limfunc{Tr}_{0}K_{0}R_{M,0}(x)R_{M-1,M}(x)\cdots
R_{1,2}(x)^{2}R_{2,3}(x)\cdots R_{M-1,M}(x)R_{M,0}(x),
\end{equation*}%
where the boundary conditions are encoded in the only non-trivial boundary
matrix
\begin{equation*}
K=\left(
\begin{array}{cc}
q^{-1} & 0 \\
0 & q%
\end{array}%
\right) \ .
\end{equation*}%
Obviously, one can widen the discussion to include the transfer matrix and
in the cases (\ref{pottsq}) and (\ref{qspecial}) the latter turns out to be
Hermitian as well.
\medskip

In comparison, different choice of products have been made in the
literature, e.g. in the context of applications to logarithmic conformal
field theory the choice $G_{ij}=\delta _{ij}$ leads to non-trivial Jordan
blocks in the Hamiltonian and transfer matrix \cite{PRZ06}. The results
presented here show that other choices might be possible where the
Hamiltonian or transfer matrix are Hermitian.
\medskip

A natural extension of our discussion is to include more complicated
boundary conditions or other representations of the Temperley-Lieb algebra %
\cite{Kulish}. Additional open problems are the formulation of the graphical
calculus for the values (\ref{pottsq}) as this would provide a more
convenient formalism for computations. The main hurdle to overcome is to
find a graphical rule for the reduction of the state space which has to be
carried out first in order to remove non-trivial Jordan blocks in the
Hamiltonian. One may also wish to extend the discussion to roots of unity
other than (\ref{pottsq}). We hope to address these questions in future work %
\cite{CK}.

{\small \medskip }

\noindent \textbf{Acknowledgments}. The author wishes to express his gratitude to the organizers of RAQUIS
(September 2007, Annecy, France) and would like to thank Philippe Di Francesco, Catharina
Stroppel and Robert Weston for discussions. C.K. is financially supported by
a University Research Fellowship of the Royal Society.

\end{document}